\documentclass[preprint]{aastex}

\usepackage{graphicx}
\usepackage{wrapfig}
\usepackage{natbib}

\newcommand{\be}{\begin{equation}}
\newcommand{\ee}{\end{equation}}
\newcommand{\nn}{\mbox{} \nonumber \\ \mbox{} }
\newcommand{\ba}{\begin{eqnarray}}
\newcommand{\ea}{\end{eqnarray}}
\newcommand{\om}{\omega}

\newcommand\eg{{\it{e.g.\ }}}

\newcommand{\Bf}{{magnetic field}}
\newcommand{\Bfs}{{magnetic fields}}
\newcommand{\Ef}{{electric  field}}

\newcommand{\NS}{neutron star}

\newcommand{\EM}{electromagnetic}

\newcommand{\ms}{magnetosphere}
\newcommand{\mss}{magnetospheres}

\newcommand{\LC}{light cylinder}
\newcommand{\Lf}{Lorentz factor}

\begin{document}


\title{Radiation formation length  in astrophysical high brightness  sources }

\author{Maxim Lyutikov,\\
 Department of Physics, Purdue University, 
 525 Northwestern Avenue,
West Lafayette, IN, USA}

\begin{abstract}
The  radiation formation length for relativistic particles, $l_c \sim \gamma^2 \lambda$ ($\gamma$ is the \Lf, $\lambda$ is  the emitted wavelength),  is  much lager than the inter-particle distances in  many astrophysical applications. This leads to the importance of plasma effects even for the high energy emission. The consequences  are nontrivial: (i) averaging of the phases of the emitting particles reduces the power (a.k.a., a circle current does not emit);   (ii) density fluctuations may lead to the  sporadic  production of coherent emission;   (iii) plasma effects during   assembly  of  a photon may lead to  the suppression of the emission (Razin-Tsytovich effect for the  superluminal modes), or,  in the opposite limit of subluminous normal modes, to the newly discussed synchrotron   super-radiance.  For synchrotron emission the  radiation formation length is  the same for all emitted waves, $\sim c/\om_B$ (non-relativistic Larmor length); for curvature emission it is  $R/\gamma$ - macroscopically long in pulsar \mss\ (\eg\ kilometers for radio). The popular  model of ``coherent  curvature emission by bunches'',  with  kilometers-long radiation formation length,  particles swinging-out  in a rotating \ms\ before they   finish emitting a wave, extreme requirements on the momentum spreads, and  demands on the electric energy needed to keep the electrostatically repulsing charges together, all  make that  model  internally inconsistent.  Long  radiation formation lengths affect how emission from PIC simulations should be  interpreted: phases of the emitted wave  should be   added over the  radiation formation length,  not just  the powers from the instantaneous acceleration of each particle.  
\end {abstract}

\section{Introduction}

The Landau-Pomeranchuk-Migdal (LPM)   effect  \citep{Ter-Mikaelyan,LPM,1956PhRv..103.1811M,1956NCim....3S.652F}, also a review by  \cite{1998PhRvD..57.3146B,1829382}, is a somewhat surprising effect (at the time of the discovery): 
highly relativistic particles, though producing short wavelength photons, do so on  a very long spacial scale, $l_c \sim \gamma^2 c/\om$. Qualitatively, {\it in the rest frame} of a particle a photon forms on scales $\sim c/\om'$ (prime denotes frequency in the rest frame). In the lab frame a relativistic electron ``catches'' up with it's own radiation: it becomes separated by a wavelength from the initial \EM\ fields on  scale $\sim l_c$. 

As we demonstrate, in astrophysical setting the  radiation formation length  for synchrotron and curvature emission $l_c$ is large, much larger than the microscopic scales like the  inter-particle distances and the  skin depth, and can be even macroscopically large, a fraction of a system size (\eg\ pulsar \ms).
Any disturbance to the particle motion on scale $\leq  l_c$ will interfere with the production of a photon. Typically this leads to a reduction of emissivity, or even complete cancellation (continuous circular current, even composed of highly relativistic particles, does not emit).  But occasionally and/or for some combination of parameters this long radiation  formation length may lead to the enhancement of the emitted frequency/power, \eg\  via the ``chirp'' effect.

\section{Astrophysical example of long radiation formation length}

\subsection{Synchrotron emission}

For a relativistic particle with \Lf\ $\gamma \gg 1$  on circular orbit  with the radius of curvature $R$ (synchrotron or curvature emission)  the  emitted frequency is $ \om \sim \gamma^3 (c/R)$.  The radiation formation length  is then  $l_c \sim  \gamma^2 c/\om  \sim R/\gamma$. 
For synchrotron emission the radiation formation length is the non-relativistic Larmor radius
\be
l_{c,s} = \gamma^2 c/\om = c/\om_B 
\ee
$\om_B =  eB/(m_e c) $. 
It is much smaller that the relativistic Larmor radius $\sim \gamma c/\om_B$. Note that the size of the radiation formation length is the same for all  emitted wavelengths, radio or gamma:  it is determined just by the value of the \Bf\ (higher energy particles emit shorter waves by $\gamma^2$,  but the radiation formation length is longer by $\gamma^2$ - two factors cancel).

For example, in  the Crab Nebula, estimating \Bf\ as  $B\sim 10^{-4}$ G (\Bf\ is typically smaller overall, \cite{2008ARA&A..46..127H}, but may become enhances in special places like the  Inner Knot \citep{2015MNRAS.454.2754Y,2016MNRAS.456..286L}), 
the corresponding radiation formation length 
\be
l_{c,Crab} = c/\om_B = 1.7 \times 10^7\, {\rm cm}
\label{lcCrab} 
\ee
It is the same for all synchrotron-emitting particles, from radio to X-rays. The length (\ref{lcCrab}) 
 is typically much larger than the inter-particle distance of radio emitting particles, $ n ^{-1/3} \approx 10^2 $ cm 
\citep{1970ApJ...159L..77S,1999A&A...346L..49A,2020ApJ...896..147L}

\subsection{Curvature emission in pulsar \mss}

For curvature emission in pulsar \mss,
to produce  radio emission at  $\nu$ in a \Bf\ with the radius of curvature $\sim R_{LC}$ (\LC\ radius), the required \Lf\ is
 \be
 \gamma_R \sim  \left( \frac{\nu } {  \Omega} \right) ^{1/3} \sim 3 \times 10^2 \nu_9 ^{1/3}
 \ee
 (frequency normalized to 1 GHz). The corresponding  radio coherence length
 \be
 l_{c,R} = \frac{c}{ \nu^{1/3} \Omega ^{2/3}} = 5 \times 10^5  \nu_9 ^{-1/3}  \, {\rm cm}
 \label{pulsar} 
 \ee
 about 5 kilometers!  
  
 These scales are orders of magnitude  larger than the inter-particle distance at the \LC
 \ba && 
 \Delta r= n^{-1/3} = \frac{1}{2\pi}  \frac{1}{\kappa^{1/3} } \left( \frac{ e c^4 P^4}{ B_{NS} R_{NS} ^3 }  \right)^{1/3} = 10^{-3} \kappa_3^{-1/3}\, {\rm cm} 
 \nn &&
 n = \kappa n_{GJ}
 \ea 
 where for numerical estimates above  we used surface \Bf\ $B_{NS}$ and period of Crab pulsar $P=0.034$ seconds ($n_{GJ}$ is the Goldreich-Julian \citep{GJ} density, $\kappa \sim 10^3$ is the multiplicity factor).

 There is another effect that completely kills any model of ``coherent curvature emission by bunches":  rotation. Large coherence scale (\ref{pulsar}) assumes particle emission along a fixed magnetic field line. During the  formation of the pulse,  over a distance of 5km, the initial waves produced by a particle will add with waves produced by completely different particle on field lines separated by many-many wavelength.  In addition, the wave-fronts become substantially  curved: this further destroys the coherence.  (See \S \ref{spread} for additional constraints on the momentum dispersion). The curvature radio emission in rotating \mss\ cannot in principle be coherent.

 Even for GeV photons, $\nu = 2 \times 10^{23}$ Hz, the radiation formation length  for curvature emission evaluates to 
 \be
 l_{c, GeV}  \approx 7  {\rm cm},
 \label{lGeV}
 \ee
 also much larger than the inter-particle distance. 
 Thus presence of plasma may  influences  even the curvature emission of GeV photons (\eg\ there will be many background particles  within  $ l_{c, GeV}$ that may modify or interfere with the production of $\gamma$-ray emission by  the curvature mechanism).

\subsection{Gamma Ray Bursts and AGN jets}

In Gamma Ray Bursts (GRBs), within the  synchrotron prompt emission model \citep{1996ApJ...473..204S,1998ApJ...497L..17S,2004RvMP...76.1143P},    estimating the isotropic luminosity as
\be
L_{iso} = \Gamma^2 B^{\prime, 2} r^2 c \approx \Gamma^2 n' m_e c^3 r^2
\ee
($\Gamma$ is the \Lf\ of the out flow, $B'$ and $n'$ are \Bfs\  and density in the outflow frame, $r$ is the emission radius),  
we find  that  the ratio of the radiation formation length to the inter-particle distance in the outflow frame $l_{c, jet}' $ is much larger than unity
\ba && 
l_{c, jet}' = \frac{m_e c^{5/2} }{e} \frac{ r \Gamma}{\sqrt{L_{iso} }}
\nn &&
n' = \frac{L_{iso}}{ m_e c^3 r^2 \Gamma^2}
\nn &&
l_{c,GRB}'  n^{\prime, 1/3} =  \frac{ m_e ^{2/3} c^{3/2}  }{e} \frac{ r^{1/3} \Gamma^{1/3} } {L_{iso} ^{1/6}}= 
7 \times 10^3 \Gamma_2 ^{1/3}  r_{14} ^{1/3} L_{iso, 51}^{-1/6} \gg 1
\ea
(note a weak dependance on the parameters.)

Similar estimate can be made for AGN jets:
\be
 l_{c,AGN}'  n^{\prime, 1/3} =   
7 \times 10^5 \Gamma_1 ^{1/3}  r_{18} ^{1/3} L_{iso, 46}^{-1/6} \gg 1
\ee

\subsection{Non-linear    Thomson scattering in FRBs}

Importantly, $l_c$ is a ``parallel'' coherence scale, when a particle emits mostly along the direction of its motion (\eg\ synchrotron or  curvature emission). 
Compton scattering by a relativistic particle will have the same radiation formation length: it will be of the order of the wavelength of the initial  low frequency photon.

Non-linear    Thomson scattering deserves special attention. This refers to the electron in non-linear 
 EM  waves with the nonlinearity parameter  $a \gg  1 $,
 \be
a \equiv \frac{e E_w}{m_e c \om}
\label{Akhiezer} 
\ee
where $E_w$ is the \Ef\ in the coherent wave, and $\om$ is the frequency (parameter $a$ is Lorentz invariant). This process can occur within the \mss\ of the FRB-producing magnetars \citep{2014ApJ...785L..26L,2016MNRAS.462..941L,2019arXiv190103260L,2021arXiv210807881B,2021arXiv211008435L}.

  In a nonlinear wave  (without guiding field)  a particle oscillates with $\gamma_\perp \sim a$. For example, in a circularly polarized  wave, {\it in the gyration frame} (denoted with prime),  a particle moves with   \Lf\ $\gamma \sim a$  along   a circular  trajectory  of radius $c/\om_w'$ ($\om_w'$ is the frequency of the wave that is been scattered, as measured in the gyration frame). Emitted frequency is $\om \sim a^3 \om_w'$, (emitted wavelength $\lambda' \sim (c/ \om_w') /a^3$). Transverse (to the  propagation of the initial wave) coherence scale is $l_r' = (c/\om_w' )/ a$. 
In addition,  the incoming EM wave accelerates a particle along the direction of wave propagate by the ponderomotive force  to $\gamma_\parallel \sim a$, so that $\om_w'= \om_w/a$.  The radiation formation length in the frame where the particle was at rest initially is then  the wavelength of the initial radio wave,  $l_r = l_r' \sim  (c/\om_w ) = a^4 \lambda$ (centimeters for the initial  radio wave, much larger than inter-particle distance).  \citep[Non-linear    Thomson scattering is suppressed by the guide field][]{2021arXiv210207010L}


\subsection{Summary: astrophysical synchrotron/curvature emission  sources}

Our estimate show a consistent picture: the synchrotron/curvature radiation formation lengths are many orders of magnitude larger than the inter-particle distance in all astrophysically important settings. In the case of radio pulsars, the coherence  length is kilometers: add rotation and curved wave front -  curvature radio waves cannot add coherently. 

 Next, in \S \ref{Two} we provide a mathematical description of the corresponding relations. \citep[In some way, they are counter-intuitive, as illustrated by the discussion between  Landau and Ter-Mikaelyan][]{Priroda,1829382}.

\section{Two particles on circular trajectory: conditions for lasing and destructive interference} 
\label{Two} 

 Let us illustrate the previous  discussion with    a  clear/simple condition for coherent addition of waves emitted by  two particles on a circular trajectory. First we consider particles with the  same energy, \S \ref{mono},  and in \S \ref{spread} consider effects of the momentum spread.

\subsection{Phase decoherence of mono-energetic beam}
\label{mono}

Single particle EM emission is described by  the well know Lenard-Wiechert potentials \citep{Jackson}.
  Independent emission by many particles is,  in a sense, a trivial extension: a sum of independently emitted wave
(independent in a sense that each particle does not react to the \EM\ waves of the other particles). 
Importantly, amplitudes of waves should be first added, before squaring to obtain intensity.

We start with a textbook formula for the vector potential produced by a particle moving along a given trajectory, Eq. 14.67 of \cite{Jackson}, that we rewrite in the form
\ba &&
\frac{d I}{d\om d\Omega}= \frac{e^2 \om^2}{4\pi^2 c} \left| {\cal A} \right|^2 
\nn &&
{\cal A} = \int dt\,  {\bf n} \times ( {\bf n} \times {\bf \beta}) e^{i \om (t-  {\bf n}\cdot {\bf r }/c)}
\ea
${\cal A} $ is just a slightly renormalized vector potential.

Consider  next {\it two}  particles moving along a circular trajectory and  separated in angle by  
\be
2 \phi_0 \ll 1/\gamma
 \label{phi0} 
 \ee
 (this condition will be shown to be self-consistent, Eq. (\ref{phi01})).
\be
{\cal A} = {\cal A}_1+{\cal A}_2 \approx 
 {\bf n} \times ( {\bf n} \times {\bf \beta}) \left( \left. e^{i \om (t-  {\bf n}\cdot {\bf r /c})}\right|_1 + \left. e^{i \om (t-  {\bf n}\cdot {\bf r }/c)}\right|_2 \right)
 \ee
 Given (\ref{phi0}), the difference in the  $ {\bf n} \times ( {\bf n} \times {\bf \beta}) $ term is of higher order.

Let time $t=0$ be when the middle point between the particle is at origin of coordinates.  
The arguments of the exponents are then
\be
\om \left( t - {\bf n} \cdot {\bf r} (t)  /c\right)  = 
\om \left( t- (R/c)  \sin ( \beta t/R \pm \phi_0 ) \cos \theta \right)
\ee

The arguments can be separated into $\phi_0$-independent $\Phi_0$, $\phi_0$-even  $\Phi_e$, and $\phi_0$-odd $\Phi_o$ terms correspondingly: 
\ba && 
\Phi_0 =\left( \frac{1}{\gamma^2} + \theta^2 + \frac{ ( c t)^2}{3 R^2}\right) \frac{\om t}{2} 
\nn  && 
\Phi_e=  \left (  1- \frac{( c t)^2}{6 R^2} \right) \sin ^2 (\phi_0/2) \theta ^2 \om  t
\nn  && 
\Phi_o =\frac{1}{2} \left(   1- \frac{( c t)^2}{2 R^2} \right)  \sin (\phi_0) R  \theta^2 \om t
\ea
where relativistic expansion $\gamma \gg 1$, $\theta \ll 1$ has been made; also condition (\ref{phi0}) implies that emission cones for two particles nearly coincide.

A sum of two waves can be factorized
\be
e^{ i ( \Phi_0+ \Phi_e+\Phi_o)}+ e^{ i ( \Phi_0+ \Phi_e-\Phi_o)}=
2 e^{ i ( \Phi_0+ \Phi_e)} \cos \Phi_o \approx 2 e^{ i  \Phi_0} \cos \Phi_o 
\ee
$\Phi_e$ is of higher order, and can be neglected, $\Phi_e \ll \Phi_0$.

In the odd component $\Phi_o$, since important integration times are $ \Delta t \sim R/\gamma$, the time dependence can be neglected (in $\Phi_o$ the term  $(c t/R)^2$ should be compared with unity, while in $\Phi_0 $ it is compared with $1/\gamma^2$,  and thus should be retained there) 
\be
\Phi_o =\frac{1}{2}  \cos  ( \theta ^2 \sin \phi_0 (\om R)/c ) \approx \frac{1}{2}  \cos  ( \theta ^2  \phi_0 \om R/c) 
\ee

Integration over times gives
\ba &&
\frac{d I}{d\om d\Omega}=4 \times \frac{e^2 }{3 \pi^2 c}  \left( \frac{\om R} {c}\right)^2 \left( \frac{1}{\gamma^2} + \theta^2 \right) ^2 
\left( K^2 _{2/3} (\xi) + \frac{\theta^2} { 1/\gamma^2 + \theta^2}  K^2 _{1/3} (\xi)  \right) \times  \cos ^2 ( \theta ^2 \om R  \phi_0/c)
\nn && 
\xi = \frac{R \omega  \left(\frac{1}{\gamma ^2}+\theta ^2\right)^{3/2}}{3 \sqrt{2}}
\label{Iomega}
\ea

The coherence condition is controlled by the term  $  \cos^2  ( \theta ^2  \phi_0 \om R/c)$.
For $\phi_0 =0$ we have  emissivity   $4 $ times  the single particle due to  the coherent  addition of waves.

 For longer separation,   waves emitted by two particles will interfere destructively when
$ \theta ^2 \phi_0  \om R / c  \sim 1$.  Estimating $\theta \sim 1/\gamma$, $\om \sim \gamma^3 c/R$, this implies
\be
\phi_0 \sim 1/\gamma
\label{phi01}
\ee
Or in physical distance $l_c \sim R/\gamma$. Thus,  in calculating emission, contribution from a  whole  region  $ \sim  R/\gamma = \gamma^2 c/\om$ should be taken into account.

\begin{figure}[h!]
\centering
\includegraphics[width=.99\textwidth]{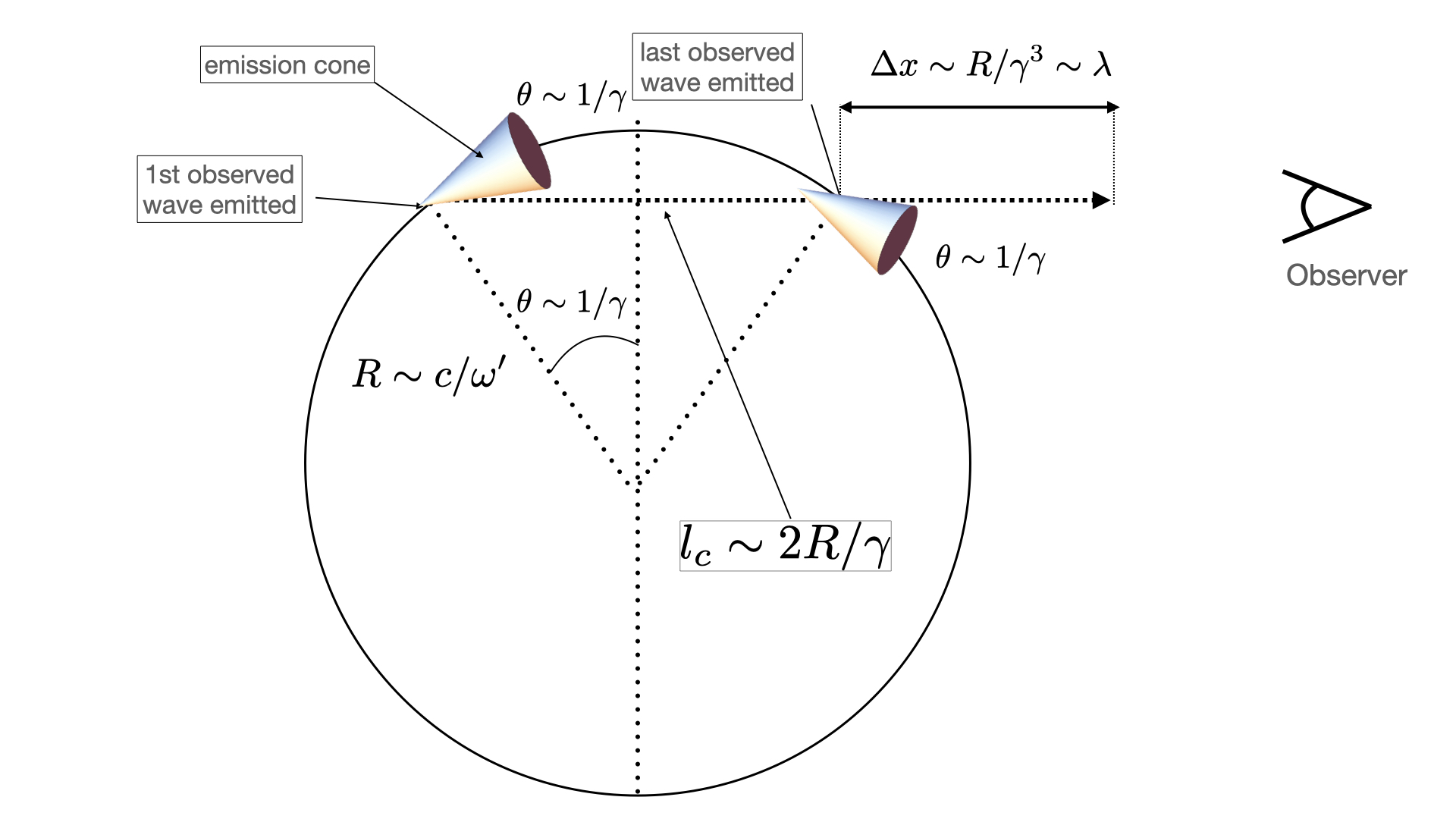} 
\caption{Illustration of long radiation formation length for a highly  relativistic particle  moving along a circular trajectory.  Particle-wave phase slippage occurs on scales $\sim R/\gamma \sim \gamma^2 \lambda$. This illustrates a transition  "single particle emits/constant current does not".}
\label{Picture-of-RadiationFormation}
\end{figure}

To see this effect qualitatively, consider a particle that starts/ends to emit towards an observer when it's direction of motion is at angle  $\sim 1/\gamma$ towards an observer    (in vacuum), 
  Fig. \ref{Picture-of-RadiationFormation}. At the end of emission the particle is behind the initial wave front by 
\be
( \Delta x) = \frac{4}{3} \frac{R}{\gamma^3}
\label{deltax} 
\ee
(inverse of this gives the typical frequency $\sim \gamma^3 c/R$.) Yet the total length a particle covered during the emission time is $l_c =2 R /\gamma$.
Particles (of the same energy) separated by distance $\ll l_c$ emit coherently. Waves emitted by  particles separated by distance $\sim l_c$  add destructively. 

\subsection{Effects of the momentum spread of the beam} 
\label{spread}

Intrinsic momentum spread of the beam will also contribute to decoherence. The corresponding conditions are severe, as we show below. Qualitatively,  for coherent addition of emitted waves a delay between two particles over the radiation formation length should be less than (\ref{deltax}). The curvature and synchrotron emission mechanism are somewhat different in this respect.
For curvature emission
\be
\Delta \left( \frac{1}{\gamma^3}  \right) \approx \frac{( \Delta \gamma)/\gamma}{\gamma^3} \leq 1
\label{deltag} 
\ee
 This implies  that a  minuscule  spread of velocities can be tolerated:
\be
(\Delta v) \sim 1/\gamma \leq 10^{-3}
\ee
(since $(\Delta \gamma)  =( \Delta v) \gamma^3$). 
Suppression of coherent curvature emission, and of the related     \cite{1971ApJ...170..463G} instability by small velocity dispersion has been pointed out in many previous papers, \eg\, \citep{1977MNRAS.179..189B,1990MNRAS.247..529A,1992RSPTA.341..105M,1999ApJ...521..351M,2021MNRAS.500.4530M}. Out derivation outlines a simple heuristic estimate. 

In  the case of curvature $\gamma$-rays,  the condition (\ref{deltag}) would prevent coherent emission, even though there may be many $\gamma$-ray emitting particles within the radiation formation zone (\ref{lGeV}).

For synchrotron emission,  in order to emit coherently two particles with different energies should  rotate around different \Bf\ lines (since $R \sim \gamma c/\om_B$ in this case). The coherence condition then becomes
\ba &&
\Delta \left( \frac{1}{\gamma^2}  \right) \approx \frac{( \Delta \gamma)/\gamma}{\gamma^2} \leq 1
 \nn &&
 (\Delta v) \sim 1
 \ea
 (since $( \Delta \gamma)/\gamma \ll 1$ particles emit approximately within the same solid angle).
 This condition is less demanding than of the curvature emission: coherent synchrotron emission (\eg\ by fluctuation of plasma density)  has a chance.

\section{Implications}

The implication of many particles within the radiation formation length   are not straightforward. A  photon has not formed yet, there is no   ``a frequency of a photon'', it's a complicated mix of phases and amplitudes  of (re)-emitted waves. A complete analysis would involve a detailed statistical plasma consideration, including fluctuation-dissipation relations
\citep{1975OISNP...1.....A,1978JETP...48...51S}; a plasma approach, even kinetic, which assumes participation of many  particles may not even be applicable to some parts/parameters  of the problem.

Another issue is a spectrum of energy of particles/presence of low energy background population: for higher energy particles the lower energy one may play a role of a "background". The ensuing discussion is a  mix of these different processes that can affect the photon generation.  .

\subsection{Lower energy (radio) particles, no background}
\label{Low} 

This is the most problematic regime:   a  collection of just emitting ``low''  energy particles (\eg\ in low frequency radio) will produce destructive interferes of the emitted waves.  
If there are  many ``active''  particles, the ones  that contribute to the production of the radiation, the destructive interference will lead to the suppression: many particles with emission length, with random phases: 
constant current does not emit. This  clearly applies to curvature emitting radio  particles in the \NS\  \mss: large coherence length, rotation, curvature of the wave front, little tolerance to energy spread, all  make the 'coherent curvature emission by bunches an internally not self-consistent model.

The case of    low-frequency emission in the PWNe (and by extension of  ANGe jets) is uncertain. At ``zeroth'' order it should not work:  destructive interference of many emitting particle should reduce the emissivity.  For low frequency emitting particles there is no background plasma:  emission phases from many particles within the radiation formation zone add (mostly) destructively.  This presents  a problem for radio emission of PWNe, a notoriously difficult problem \citep{kennel_84}.

A possible resolution is that   density fluctuations occasionally lead to the constructive addition of waves, creation of coherent emission which would  dominates the observed spectrum.
The emission will be controlled by (\ref{Iomega}) 
\be
\left| \sum_i  e^{ \theta ^2 \om R  \phi_i/c} \right| ^2 
\ee
where the sum is over locations  $\phi_i$ of all particles contributing to the pulse. (A fluctuating excess of particles $( \delta n) \sim \sqrt{n}$, $n$ is the average number of particles within the coherence volume, emitting coherently, $\propto (\delta n)^2$,  on average would result in the same power as all particles emitting non-coherently.)


\subsection{Higher energy (X-ray -- $\gamma$-ray) particles: effects of background plasma}
\label{middle} 
  
  In this regime a high energy (X-ray -- $\gamma$-ray emitting particle propagates through a background of passive plasma. The  background plasma still  modifies  the  emission of high energy particles.  A forming photon can still be expanded in Fourier frequencies.  Different frequencies will be affected differently by the presence of the  dispersive background plasma.

 First,   low frequencies (scales longer than the plasma skin depth,  $\om \leq \om_p$, $\lambda \geq c/\om_p$ are cut out completely
 \footnote{Below, for clarity,  in most relations we  neglect relativistic effects; it is understood that many relations will be modified by the relativistic effects in the background plasma, \eg\ (\ref{gammaCHgammap}).}. But these frequencies are too low to be of interest anyway.
 
 Second,   for somewhat higher ``middle'' frequencies, one can treat the \EM\ signal from the higher energy  emitting particle as a set of waves propagating through a background plasma. 
  The background plasma will modify waves' dispersion (possibly in a frequency-dependent manner depending on  the dispersion in the bulk plasma). This will affect   the final coherent addition of phases. This may even lead to amplification of the radiated power if compared with singe particle emissivity, see   \S \ref{middle}.
     This type of interaction requires that the corresponding interaction can be treated in a continuous limit - far from a sure condition.
 
Finally, the highest frequencies with wavelength shorter than the inter-particle distance  suffer from single-particle scattering. These processes are typically not important since the scattering depth is too low, but may become important in the nonlinear regime $a\gg 1$ \citep{2004PlPhR..30..196B,2019arXiv190103260L,2021arXiv210807881B}.
 
\subsection{Plasma effects on higher energy particles}
\label{effects}

The ``middle'' regime, 
    is the most interesting case: how changing dispersion of waves been emitted by the  higher energy particles affects the final properties of a photon. Let us consider it in more detail.

The corresponding  modifications depend on the  dispersion relations of  the normal modes. Two related, but somewhat different effects are at play:  changing emission cone (from $\sim 1/\gamma$), and wave-particle catching-up conditions. A combination of the two affects the  observed pulse duration (and hence typical emission frequency).

 For superluminal modes (\eg\ waves in unmagnetized plasma with phase velocity $v_{\rm ph} = c (1+ \om_p^2/(2 \om^2))$ the emission  synchrotron  cone increases, 
\be
\phi_{em} \sim  \sqrt{ \frac{1}{\gamma^2} + \frac{\om_p^2}{\om^2}}
\label{phiX}
\ee
This leads to the suppression of low frequencies synchrotron emission  \citep[Razin-Tsytovich effect,][]{1965ARA&A...3..297G}  for 
\be
\om \leq {\gamma} {\om_p} = \frac{\om_p^2}{\om_B}
\label{RT}
\ee

In magnetically dominated plasma with
\be
\sigma = \frac{\om_B^2}{\om_p^2} \geq 1
\ee
 modes are subluminal, 
\be
v_{\rm ph} = c \left(1-  \frac{\om_p^2}{2 \om_B^2}\right)
\ee
for $\om_B \gg \om, \,\om_p$. (This is the regime in 
 pulsar \mss, and might be applicable to some regions in PWNs,  Eq.  (\ref{sigmaPWN})).

This leads to the decrease of the emission cone,
\be
\phi_{em} \sim  \sqrt{ \frac{1}{\gamma^2} - \frac{\om_p^2}{\om_B^2}}
\ee
Condition $\phi_{em}=0$ in this case corresponds to the transition to the Cherenkov-type emission \citep[and magnetically induced emission at the anomalous Doppler effect,][]{1970pewp.book.....G,mu79,Kaz91,1999ApJ...512..804L,1999MNRAS.305..338L}

The case of subluminal background modes leads to the  effect of super-radiance, that deserves a special attention, \S \ref{superradiance}. 

 \subsection{Synchrotron/curvature   super-radiance}
\label{superradiance} 

The ``middle''  regime, \S  \ref{middle}-\ref{effects}, offers an interesting possibility of super-radiance.
Consider  synchrotron emission of fast particles in magnetically-dominated case, $\om_B \geq \om_p$ (properly defined, taking possible \Lf\ of random motion into account).  Consider a high energy particle in cyclotron motion around \Bf. The particle propagates perpendicular to the field. There are two modes: O-mode (with polarization along the \Bf) with phase velocity $v_{ph, O} \approx c$, and X-mode (with polarization perpendicular to  the \Bf) with phase velocity 
\be
v_{ph, X} \approx c \left( 1- \frac{\om_p^2}{2 \om_B^2} \right)
\ee
for $\om \leq \om_B$ \citep{AronsBarnard86,1999JPlPh..62...65L}. 

In the case of curvature  emission, waves  propagating along the \Bf\ are of X-mode type. Below we analyze these two cases in parallel.

Let a particle move with $v\leq v_{ph, X}$, in the normal Doppler effect region, but close to the  Cherenkov condition, 
\ba &&
\gamma \approx  \frac{ \om_B} {\om_p} - (\delta \gamma) = \gamma_{Ch}  - (\delta \gamma)
\nn &&
\gamma_{Ch} = \frac{ \om_B} {\om_p} 
\ea
The emission cone for the X-mode (\ref{phiX})  is then
\be
\phi_{em,X} \approx \sqrt{2} \left( \frac{\om_p}{2 \om_B} \right)^{3/2} \sqrt{(\delta \gamma)} \approx \gamma_{Ch}^{3/2}  \sqrt{(\delta \gamma)} 
\ee

The wave-particle delay between on-off moments is 
\be
\Delta x = \frac{ 8 \sqrt{2}}{3}  \left(  \frac{ \om_p} {\om_B} \right) ^{9/2}  (\delta \gamma)^{3/2} R\approx
 \gamma_{Ch}^{9/2}  (\delta \gamma)^{3/2} R
\ee
Typical emitted frequency
\ba &&
\om \sim \frac{c}{\Delta x} = \frac{3} { 8 \sqrt{2}} \left(  \frac {\om_B}  { \om_p}\right) ^{9/2}  (\delta \gamma)^{-3/2}  \frac{c}{R} \approx
\frac{\gamma_{Ch}^{9/2}}  { (\delta \gamma)^{3/2} }\frac{c}{R}  
=
\nn && 
 \frac{3} { 8 \sqrt{2}} \left(  \frac {\om_B}  { \om_p}\right) ^{7/2}  (\delta \gamma)^{-3/2} \om_B =
\frac{  \gamma_{Ch}^{7/2} }{ (\delta \gamma)^{3/2} } \om_B
\label{typicaomega}
\ea
Divergence at $(\delta \gamma) \to 0$ corresponds to the transition to the Cherenkov-type emission. A particle arbitrarily close to the Cherenkov resonance in a medium  with $n_r \geq 1$ can produce arbitrarily high frequency (of course, given the limitations of the regions of $n_r \geq 1$).

Radiation formation length is:
\ba&& 
l_r=  4 \sqrt{2} \left(  \frac{ \om_p} {\om_B} \right) ^{3/2}  (\delta \gamma)^{1/2} R
\approx  \frac{ (\delta \gamma)^{1/2} }{\gamma_{Ch}^{3/2} }R 
 \to 
 4 \sqrt{2} \left(  \frac{ \om_p} {\om_B} \right) ^{1/2}  (\delta \gamma)^{1/2}  \frac{c}{\om_B} \approx \sqrt{ \frac{ (\delta \gamma)} {\gamma_{Ch}}} \frac{c}{\om_B}
 \nn &&
 \frac{l_r}{c/\om} \approx  \left(  \frac {\om_B} { \om_p} \right) ^{3} (\delta \gamma)^{-1}  = \gamma_{Ch}^3(\delta \gamma)^{-1} \gg 1
 \label{lecH}
 \ea
 Recall that it is required that the radiation formation length is sufficiently large, so that there are many background particles with the zone to ensure modification of the waves' suppression. The coherence length  (\ref{lecH})  is small, but it must still satisfy the condition of continuous treatment of the background plasma.  The above relations, when expressed in terms of $\gamma_{Ch}$ and $R$ are also applicable to the curvature emission. 

Peak spectral power and total power
\ba &&
P_\om \approx \frac{e^2}{c} \left(  \frac{\om_B}{ \om_p}  \right) ^{1/2} \frac{ \om_B}{(\delta \gamma)^{1/2} }
\nn &&
P_{tot} \sim P_\om \om \approx  \frac{e^2}{c}   \frac{\om_B^6}{ (\delta \gamma)^2 \om_p^4} =  \frac{e^2}{c} \frac{\gamma_{Ch}^4}{ (\delta \gamma)^2 }\om_B^2 
\ea
This is much larger than the peak  spectral power of a particle with \Lf\ $\gamma_{Ch} $ in vacuum,
\ba && 
\frac{P_{tot} }{P_{vac} } \approx \frac{\gamma_{Ch}^2}{ (\delta \gamma)^2 }
\nn &&
P_{vac}  \approx  \frac{e^2}{c}  \gamma_{Ch}^2 \om_B^2 
\ea

Qualitatively, the mechanism of synchrotron   super-radiance is reminiscent of the final signal assembly  stage  in  Chirped Pulse Amplification  schemes in laser physics \citep{1985OptCo..56..219S}.

Effects of  synchrotron super-radiance may be important  in pulsars and in pulsar wind nebula.
In pulsar \mss\  it is important  to take account for the \Lf\ of the bulk plasma in calculating the Cherenkov condition \citep{1999JPlPh..62...65L,1999ApJ...512..804L}:
\be
\gamma_{Ch}= \gamma_p^{3/2} \frac {\om_B}{\om_p}  = 2\times 10^5 \gamma_{p,2}^{3/2}  \kappa_{3}^{-1/2} 
\label{gammaCHgammap}
\ee
This is approximately two orders of magnitude less that is required to produce vacuum-type GeV emission,
$\gamma \sim (R_{LC} \epsilon_{GeV} / (\hbar c))^{1/3} =2  \times 10^7$. 
 
 The typical emission energy (\ref{typicaomega}) evaluates to 
 \be
\epsilon \approx 10^{7} \, {\rm MeV}  \left(  \frac{\gamma_{Ch}} {(\delta \gamma) } \right) ^{-3/2} 
 \ee 
 Strong dependance on $(\delta \gamma)$ indicates that much higher frequencies are produced by particles close to the Cherenkov condition. Since the emission conditions are divergent near $\gamma_{Ch}$, the numerical estimates are very sensitive to the assumed parameters.

In the case of PWNe, 
plasma within (some regions) of  the Crab Nebula might be mildly magnetically dominated, 
\be
\sigma = \frac{B^2}{4\pi \gamma m_e n c^2} \approx 10 B_{-4}^2 n_{-6} ^{-1} \gamma_{2} ^{-1}
\label{sigmaPWN}
\ee
This implies that normal modes are subluminous.

\section{Discussion}

The radiation formation scales $l_c$  in high energy astrophysical sources can be much larger than the microscopic scales (wavelength, skin/Debye scale), and actually can be
 macroscopically long  (a fraction of the system size  for curvature radio emission in \NS\ \mss).
 For synchrotron emission it  is the non-relativistic Larmor scale, $c/\om_B$, for curvature it is  $R/\gamma$.

We come to a surprising conclusion: in nearly all astrophysical situations there are many particles within the radiation formation length for synchrotron and curvature emission.  
Any perturbation of a  particle  motion, or of the accompanying the EM signal,  on  scales  shorter than $l_c$  interferes with the production of a photon.  The implies that correlations of phases of emitting particles, on macroscopic scales are important. 
This leads to the modifications of the radiation properties, when compared to vacuum case. This may lead  to suppression of radiation, but  in certain cases to amplification. 

Our  estimate demonstrate, most surprisingly,  that  collective effects (of the background plasma)   must be taken into account when calculating even the properties of $\gamma$-ray emission.
 In the extreme case,  even relativistic particles  do not produce any radiation:  ``constant circular current does not emit". In the other case, plasma effects may lead to synchrotron/curvature   super-radiance.

Various plasma effects due to possible presence of background plasma may modify the emission. 
These plasma modifications of the emission cannot be simply quantified: there are many different regime. Beside the obvious suppression due to many particles emitting out of phase,  the  background plasma can  lead to the enhancement of the radiation, even at high energies).  Also,   density   fluctuations of emitting particles on  long coherence  scales $\ll l_c$ may lead to the  occasional  production of coherent emission (even at high energies!).

Coherent curvature emission in radio, as a mechanism of pulsar radio emission, is hopeless, though: radiation formation length is  several kilometers; in the rotating \ms\ the early  emitted  waves cannot even resonate with the same particle; minuscule momentum spread destroys coherence;  also  electrostatic energetic demands on creation of bunches of repulsive charges are prohibitively high \citep{2021ApJ...918L..11L}.

There is  a clear example, where synchrotron  radio emission is observed, even though the radiation formation length is many orders of magnitude larger than the inter-particle distance, and no background plasma is expected: PWNe. The emission should then occur due to the  fluctuations of density of emitting particles with the radiation formation length. Presence of many particles suppresses emission, but occasional fluctuation will produce coherent emission, with the similar total power to the non-coherent emission.

Finally, long radiation reaction lengths may affect how results of PIC simulations are interpreted. Codes with radiation reaction force \citep[\eg][]{2013ApJ...770..147C} use instantaneous acceleration to calculate radiation losses. Effectively, this  procedure assumes very short radiation formation length. 

  This work had been supported by 
NASA grants 80NSSC17K0757 and 80NSSC20K0910,   NSF grants 10001562 and 10001521.

I would like to thank Mikhail Medvedev for discussions. 

\bibliographystyle{apj}

  \bibliography{/Users/maxim/Home/Research/BibTex}

\end{document}